\documentclass{article}
\usepackage{amssymb}
\usepackage{bm}
\usepackage[dvips]{graphicx}
\begin{document}

\title{Tunnel magnetoresistance of magnetic junctions\\ with cubic symmetry
of the layers}

\author{S. G. Chigarev, E. M. Epshtein\thanks{E-mail: eme253@ms.ire.rssi.ru}, P. E. Zilberman\\
V.A. Kotelnikov Institute of Radio Engineering and Electronics\\
    of the Russian Academy of Sciences, 141190 Fryazino, Russia}

\maketitle

\abstract{A tunnel magnetic junction is considered with magnetic hard and
magnetic soft layers of cubic symmetry. The magnetic switching is analyzed
of the layers by a magnetic field perpendicular to the initial
magnetizations. In such a situation, an additional peak of the TMR ratio appears
at the magnetic field value lower substantially than the anisotropy energy of the
soft layer.}

\section{Introduction}\label{section1}
The tunnel magnetoresistance (TMR) is one of the most important effects in
modern spintronics. It attracts interest as a beautiful phenomenon as well
as a basis of promising spintronic devices such as magnetic
random access memory (MRAM), magnetic sensors, and magnetotunnel
transistors (see, e.g.,~\cite{Prinz,Fert}).

Recently, interest has revived to the magnetic junctions with magnetic
layers of the cubic-symmetrical materials, such as well-known
iron~\cite{Grabowski,Lehndorff,Wang}. Thin iron films of (001) orientation
have two in-plane easy axes, [100] and [010], with equal anisotropy
energy. This fact allows switching magnetization by magnetic field between different
easy axes, so that a possibility occurs of creating a memory with more
than two stable states, that may be laid to the basis of many-valued logic
devices. Moreover, very high TMR ratio, from 170\% at room temperature to 318\% at 10 K, has
been observed in epitaxial Fe/MgO/Fe magnetic tunnel
junctions~\cite{Wang} (note, that such high TMR ratios were predicted
theoretically~\cite{Butler,Mathon}). All this makes such structures to be very attractive
objects of study.

In this work, we consider a magnetic junction with magnetic hard and
magnetic soft layers of iron-type cubic symmetry. Switching such systems by external
magnetic field was studied under various relative magnetic configurations of
the layers~\cite{Lehndorff,Leonov}. Here, a different situation is
analyzed, when the magnetic field is applied perpendicularly, rather than
collinearly, as usual, to the initial collinear
magnetization directions of the layers. It is shown, that a new
$\Pi$-shaped resistance peak occurs under these conditions at the magnetic
field lower substantially than the anisotropy field, while the peak height
is lower significantly than the resistance peaks
corresponding to the conventional collinear configurations. Therefore, the
system in study obeys more than two resistive states switchable by an applied
magnetic field.

In Sec.~\ref{section2} we describe the structure in study and
introduce a phenomenological model. In Sec.~\ref{section3} the layer
magnetization directions are calculated as functions of the magnetic
field perpendicular to the initial directions. In Sec.~\ref{section4} the
results obtained are used to calculate the corresponding resistances. In
Sec.~\ref{section5} the results are discussed.

\section{The model and main equations}\label{section2}
Let us consider a magnetic tunnel junction with magnetic hard (1) and
magnetic soft (2) layers of iron-type cubic symmetry. The different values
of the magnetic anisotropy of the layers may be ensured with larger
thickness or pinning of layer 1. The layers have (001) orientation with
in-plane easy axes [100] and [010]. The layer magnetization vectors are
directed along the [100] easy axis originally, such parallel configuration
corresponds to minimal resistance $R_P$ of the junction when the current
flows perpendicular to the layer planes (CPP mode). Under magnetization
switching by applied magnetic field, the junction resistance changes; this is well-known TMR
effect.

The conductance of the tunnel junction is~\cite{Utsumi}
\begin{equation}\label{1}
  G(\chi)=G_P\cos^2\frac{\chi}{2}+G_{AP}\sin^2\frac{\chi}{2},
\end{equation}
where $\chi$ is the angle between the layer magnetization
vectors, $G_P$ and $G_{AP}$ are the conductances corresponding to parallel ($\chi=0^\circ$) and
antiparallel ($\chi=180^\circ$) configurations of the junction,
respectively. The angular dependence of the TMR ratio $F(\chi)$ defined as
\begin{equation}\label{2}
  F(\chi)=\frac{R(\chi)-R_P}{R_P},\quad R(\chi)\equiv\frac{1}{G(\chi)}
\end{equation}
takes the form
\begin{equation}\label{3}
  F(\chi)=\frac{f(1-\cos\chi)}{2+f(1+cos\chi)},
\end{equation}
where $f=F(180^\circ)=(R_{AP}-R_P)/R_P$ coincides with the conventional
definition of the TMR ratio~\cite{Guilliere}.

To obtain the TMR ratio as a function of the external magnetic field, it is
necessary to find the dependence of the $\chi$ angle on the value and
direction of the magnetic field.

The magnetization orientation of the magnetic layer is determined with the
minimum condition of the magnetic energy, which takes the following form
under cubic symmetry~\cite{Buschow}:
\begin{equation}\label{4}
  U(\theta)=-MH\cos(\theta-\alpha)+\frac{1}{2}MH_a\sin^2\theta\cos^2\theta,
\end{equation}
where $M$ is the saturation magnetization, $H$ is the external magnetic
field, $H_a$ is the anisotropy field, $\alpha$ and $\theta$ angles
determine directions of the external magnetic field and magnetization
vector, respectively; the angles are counted off from [100] axis.

Let the magnetic field be applied along [010] axis ($\alpha=90^\circ$).
The extremum condition $dU(\theta)/d\theta=0$ takes the form
\begin{equation}\label{5}
  [(1-2\sin^2\theta)\sin\theta-h]\cos\theta=0,
\end{equation}
where $h=H/H_a$ is the dimensionless magnetic field.

The roots $\theta=90^\circ\equiv\theta_1$ and $\theta=-90^\circ\equiv\theta_2$ of
Eq.~(\ref{5}) correspond to two equilibrium positions of the magnetization
vector; with $\alpha=90^\circ$, the latter state becomes unstable ($d^2U(\theta)/d\theta^2<0$) at
$h>1$.

The other equilibrium states are determined by the cubic equation with
respect to $\sin\theta$
\begin{equation}\label{6}
  2\sin^3\theta-\sin\theta+h=0.
\end{equation}
At $h<\sqrt6/9\equiv h_0\approx0.272$, the equation has three real roots, of which only
one corresponds to a stable state, namely,
\begin{equation}\label{7}
  \theta=\arcsin\left(\sqrt\frac{2}{3}\cos\left(60^\circ
  +\frac{1}{3}\arccos\frac{h}{h_0}\right)\right)\equiv\theta_3.
\end{equation}
At $h>h_0$, the stable solution $\theta_3$ disappears (the only real root of Eq.~(\ref{6}) corresponds to
an unstable state), and only $\theta_1=90^\circ$ and $\theta_2=-90^\circ$ stable states remain.

These results are used below to analyze magnetic configuration of the
magnetic junction.

\section{Layer magnetization direction}\label{section3}
Let us trace behavior of the magnetization vector of a magnetic layer
oriented originally along [100] easy axis in the magnetic field directed along
another easy axis [010]. With the field increasing from zero, the
magnetization vector deflects gradually from the former easy axis to the
latter one in accordance with Eq.~(\ref{7}), until $h$ reaches $h_0$
value (the corresponding value of $\theta$ angle is
$\theta=\arcsin(1/\sqrt6)\approx24^\circ$) and $\theta_3$ state disappears.
Then the vector turns abruptly to $\theta_1$ position (along the magnetic
field and [010] easy axis) and remains in this state with further
increase of the magnetic field. It remains in that state, also, when the
magnetic field decreases to zero and further to negative value $h=-1$.
Then the magnetization vector switches abruptly from the $\theta_1$
position antiparallel to the magnetic field to the parallel one $\theta_2$
position. Under subsequent cycling of the magnetic field between $h=-1$
and $h=+1$ values, the magnetization switches abruptly between $\theta_1$ and
$\theta_2$ states without returning to the initial state $\theta=0^\circ$. The
return may be realized by the magnetic field $h=h_0$ along [100] axis.

\begin{figure}
\includegraphics{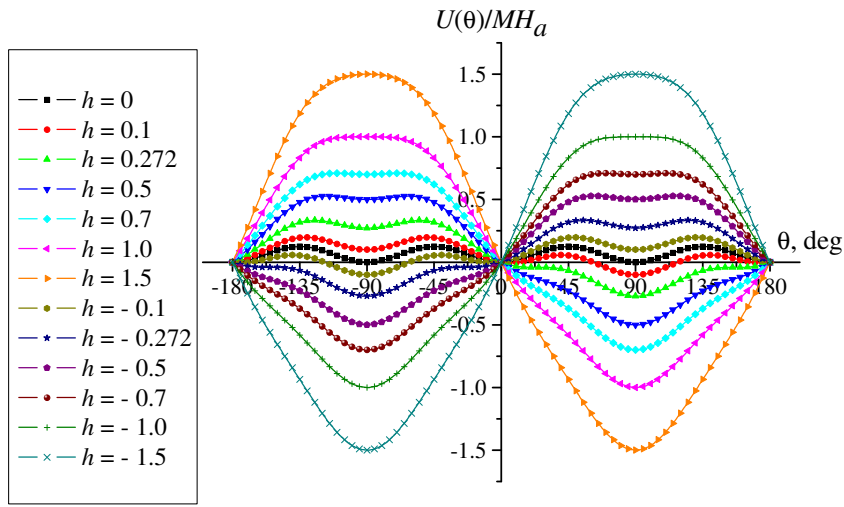}
\caption{The (dimensionless) magnetic energy of a magnetic layer with cubic
symmetry as a function of $\theta$ angle
between the magnetization vector and the easy axis at various dimensionless magnetic
fields $h=H/H_a$ directed along [010] axis ($\theta=90^\circ$). The
minimum $\theta=0^\circ$ disappears at $h=0.272$ so that the system being
in that minimum comes to another minimum $\theta=90^\circ$. The latter
exists under subsequent field changes until $h=-1$, when the system
switches to $\theta=-90^\circ$ minimum. Under subsequent cycling the
magnetic field between $h=-1$ and $h=1$, the system switches between
$\theta=-90^\circ$ and $\theta=90^\circ$ positions without returning to
the initial $\theta=0^\circ$ position.}\label{fig1}
\end{figure}

Such a behavior of the system with transitions from
one energy minimum to another one may be traced also by means of Fig.~\ref{fig1} where
the magnetic energy $U(\theta)$ is shown in various magnetic fields.

Now we consider the magnetic junction described in Sec.~\ref{section2}
instead of a single magnetic layer. The each of two layers behaves
independently of the other, so that the magnetization direction dependence on the
applied magnetic field is to be found as above for both layers, but with
different anisotropy energies. We assume, as an example, that the anisotropy
energy of the magnetic hard layer is twice as many as that of the soft layer.

\begin{figure}
\includegraphics{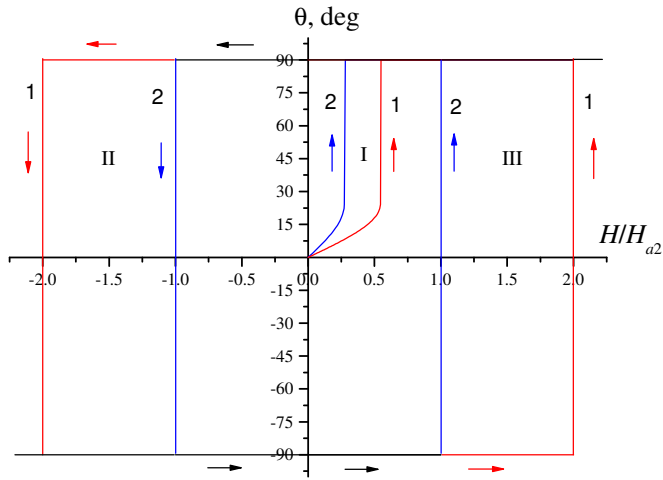}
\caption{The dependence of the magnetic hard (1) and soft (2) layer
orientations on the magnetic field along [010] axis
($\theta=90^\circ$) referred to the soft layer anisotropy field $H_{a2}$.
The layer anisotropy field ratio $H_{a1}/H_{a2}$ is supposed to be equal
to 2, as example. Three subsequent stages of the whole
switching cycle, $0^\circ\to90^\circ$, $90^\circ\to-90^\circ$, and
$-90^\circ\to90^\circ$, are marked with Roman numerals I, II and III,
respectively. The arrows show directions of the layer magnetization
switching.}\label{fig2}
\end{figure}

Because of larger anisotropy energy, the hard layer lags behind the soft
one and switches from one position to another at larger magnetic field.
The results are shown in Fig.~\ref{fig2}. Three subsequent stages of the
whole switching cycle, $0^\circ\to90^\circ$, $90^\circ\to-90^\circ$, and
$-90^\circ\to90^\circ$, are marked with Roman numerals I, II and III,
respectively, the Arabic numerals 1 and 2 refer to the hard and soft
layers, respectively.

\section{Tunnel magnetoresistance}\label{section4}
\begin{figure}
\includegraphics{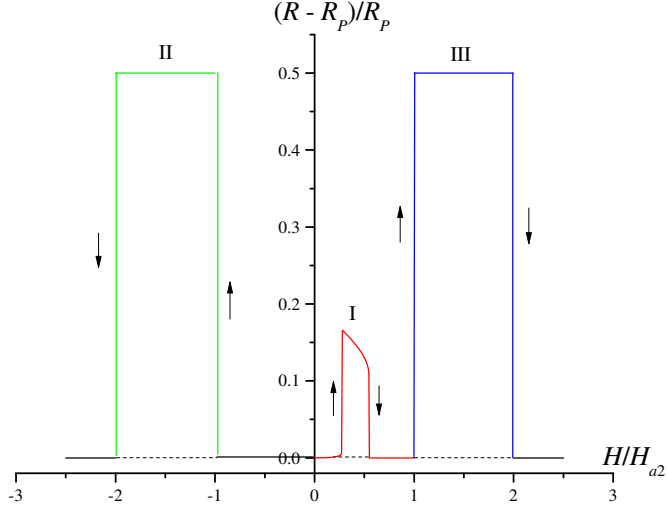}
\caption{The TMR ratio as a function of the magnetic field along [010] axis
($\theta=90^\circ$) referred to the soft layer anisotropy field $H_{a2}$ at $H_{a1}/H_{a2}=2$,
$(R_{AP}-R_P)/R_P=0.5$. The Roman numerals and the arrows have the same meaning as in
Fig.~\ref{fig2}.}\label{fig3}
\end{figure}

The TMR ratio is determined with the difference $\chi$
of the corresponding $\theta$ angles for two layers in accordance with
Eq.~(\ref{3}). Using Eqs.~(\ref{3}),~(\ref{7}), we obtain the results shown
in Fig.~\ref{fig3}. The Roman numerals have the same meaning as in
Fig.~\ref{fig2}. It is seen, that a new peak I appears besides the standard
TMR ratio peaks II and III corresponding to the switching from parallel
configuration to antiparallel one and vise versa. This new peak is lower
substantially because it corresponds to the angle between the layer
magnetization vectors smaller than $90^\circ$ (this angle tends to
$90^\circ$ when the hard layer anisotropy energy is large in comparison
with that of the soft layer; in that case, the TMR ratio is half of
the standard value). This peak corresponds to the magnetic field value
lower considerably (almost 4 times) than the soft layer anisotropy field.

\section{Conclusion}\label{section5}
It follows from the results that using magnetic junctions with cubic rather than uniaxial symmetry of
the layers opens additional possibilities in applications of the TMR
effect. The switching with a perpendicular magnetic field allows to lower
significantly the corresponding magnetic field, that may improve
sensibility of the magnetic sensors based on TMR. The extra peak of the TMR
ratio as a function of the applied magnetic field means existence of an
additional stable equilibrium state of the system in study. This fact may
be used to create memory devices with more than two stable states and
multi-valued logic devices.

\section*{Acknowledgment}
The work was supported by the Russian Foundation for Basic Research, Grant
No.~08-07-00290.


\begin{thebibliography}{12}
\bibitem{Prinz}
G.A. Prinz, \emph{J. Magn. Magn. Mater.} \textbf{200}, 57 (1999)
\bibitem{Fert}
A. Fert, \emph{Rev. Mod. Phys.} \textbf{80}, 1517 (2008)
\bibitem{Grabowski}
J. Grabowski, M. Przybylski, M. Nyvlt, J. Kirschner, \emph{J. Appl. Phys.}
\textbf{104}, 113905 (2008)
\bibitem{Lehndorff}
R. Lehndorff, M. Buchmeier, D.E. B\"urger, A. Kakay, R. Hertel, C.M.
Schneider, \emph{Phys. Rev. B} \textbf{76}, 214420 (2007)
\bibitem{Wang}
S.G. Wang, R.C.C. Ward, G.X. Du, X.F. Han, C. Wang, A. Kohn, \emph{Phys. Rev. B} \textbf{78}, 180411 (2008)
\bibitem{Butler}
W.H. Butler, X.-G. Zhang, T.C. Schulthess, J.M. MacLaren, \emph{Phys. Rev. B}
\textbf{63}, 054416 (2001)
\bibitem{Mathon}
J. Mathon, A.Umerski, \emph{Phys. Rev. B} \textbf{63}, 220403 (2001)
\bibitem{Leonov}
A.A. Leonov, U.K. R\"o\ss ler, A.N. Bogdanov, \emph{J. Appl. Phys.} \textbf{104},
084304 (2008)
\bibitem{Utsumi}
Y. Utsumi, Y. Shimizu, H. Miyazaki, \emph{J. Phys. Soc. Japan} \textbf{68}, 3444 (1999)
\bibitem{Guilliere}
M. Guilliere, \emph{Phys. Lett. A} \textbf{54}, 225 (1975)
\bibitem{Buschow}
K.H.J. Buschow, F.R. De Boer, \emph{Physics of Magnetism and Magnetic
Materials}
(Kluwer Academic Publ., New York, 2003)
\end{thebibliography}
\end{document}